\documentstyle[12pt,fleqn]{article}
\renewcommand{\baselinestretch}{1.2}

\def\npb#1#2#3{    {\it Nucl. Phys. }{\bf B #1} (19#2) #3}
\def\plb#1#2#3{    {\it Phys. Lett. }{\bf B #1} (19#2) #3}
\def\prd#1#2#3{    {\it Phys. Rev. }{\bf D #1} (19#2) #3}

\def\prl#1#2#3{    {\it Phys. Rev. Lett. }{\bf #1} (19#2) #3}

\def\mpla#1#2#3{   {\it Mod. Phys. Lett. }{\bf A #1} (19#2) #3}

\def\ibid#1#2#3{   {\it ibid. }{\bf #1} (19#2) #3}

\def\eq#1{{eq.~(\ref{#1})}}
\def\eqs#1#2{{eqs.~(\ref{#1})--(\ref{#2})}}
 \let\vev\VEV
\def\abs#1{\left| #1\right|}
\def\mod#1{\abs{#1}}
\def\Im{\mathop{\mbox{Im}}}
\def\Re{\mathop{\mbox{Re}}}
\def\Tr{\mathop{\mbox{Tr}}}
\def\etal{{\it et al.}}

\newcommand{\bea}{\begin{eqnarray}}
\newcommand{\beq}{\begin{equation}}
\newcommand{\eea}{\end{eqnarray}}
\newcommand{\eeq}{\end{equation}}
\newcommand{\nnu}{\nonumber}

\newcommand{\spav}[1]{\parbox{1mm}{\vspace*{#1}}}

\begin{document}

\begin{titlepage}
\begin{flushright}
CERN-TH.7097/93\\
SISSA 174/93/EP
\end{flushright}
\spav{0.0cm}
\begin{center}
{\Large\bf The Relevance}\\
{\Large\bf  of the  Dipole Penguin Operators in
$\epsilon '/\epsilon$}\\
\spav{1.5cm}\\
{\large Stefano Bertolini}
\spav{0.7cm}\\
{\em INFN, Sez. di Trieste, c/o SISSA}\\
{\em Via Beirut 4, I-34014 Trieste, Italy}\\
\spav{1.0cm}\\
 {\large Marco Fabbrichesi}
\spav{0.7cm}\\
{\em CERN, Theory Division}\\
{\em CH-1211 Geneva 23, Switzerland}\\
\spav{1.0cm}\\
{\large Emidio Gabrielli}
\spav{0.7cm}\\
{\em Dip. di Fisica, Universit\`{a} di Roma I ``La Sapienza''}\\
{\em P.le A. Moro 2, I-00185 Rome, Italy}\\
\spav{1.5cm}\\
{\sc Abstract}
\end{center}
The standard model contribution to $\epsilon'/\epsilon$ of the
magnetic- and electric-dipole penguin operators
$Q_{11} = \frac{g_s}{16\pi^2} \, m_s \: \bar{s}\, \sigma_{\mu\nu}
 t^a G^{\mu\nu}_a (1-\gamma_5 )\, d $ and
$Q_{12} = \frac{eQ_d}{16\pi^2}\, m_s \: \bar{s}\, \sigma_{\mu\nu}
 F^{\mu\nu} (1-\gamma_5 )\, d $ is discussed.
While the electromagnetic penguin $Q_{12}$
seems to have a vanishingly small
matrix element, we find that the gluonic dipole operator $Q_{11}$
may give a contribution to $\epsilon'/\epsilon$
comparable to that of other
operators so far considered, and should therefore be consistently
included in the analysis.

 \vfill
\spav{.5cm}\\
CERN-TH.7097/93\\
SISSA 174/93/EP\\
 December 1993, revised March 1994.

\end{titlepage}

\newpage
\setcounter{footnote}{0}
\setcounter{page}{1}

{\bf 1.} The determination of the $\Delta S =1$,
$CP$-violating parameter $\epsilon '/\epsilon$ in neutral kaon decays has
attracted considerable experimental and
theoretical interest (for a complete list
of references see, for instance, ref.~\cite{epsilon}).
The possibility
of drastic cancellations among QCD- and electroweak-induced
operators~\cite{Lisa}, occurring
for large values of the top quark mass,
has spurred new and more
accurate theoretical investigations~\cite{Buras,Roma}.
On the experimental side, the present sensitivity
of 1 part in $10^{3}$~\cite{experiments} is not conclusive  and an effort in
improving it by one order of magnitude is under way.

The theoretical framework for the study of $\epsilon'/\epsilon$
is the effective field theory.
The local
Hamiltonian for $\Delta S = 1$ transitions can be written,
for $\mu<m_c$, as~\cite{Buras}
\beq {\cal H} = \frac{G_F}{\sqrt{2}} \lambda_u  \sum_i \Bigl[
z_i(\mu) + \tau y_i(\mu) \Bigr] Q_i (\mu)
 \, . \label{ham}
 \eeq
The list of the effective operators $Q_{i}$ ($i=1-10$) is reported
in refs.~\cite{Buras,Roma}, whose notation
we  follow closely and where the reader may find a complete
discussion of the basic tools we use in the present analysis.
We just recall that
$Q_{1,2}$ stand for the ``tree-level'' $W$-induced current-current
operators, $Q_{3-6}$ for the
QCD penguin operators and $Q_{7-10}$ for the electroweak penguin (and box)
ones.

The Wilson coefficients $z_{1,2}(\mu)$
run from $m_W$ to $m_c$ via the corresponding $2\times 2$ sub-block
of the $10\times 10$ anomalous dimension matrix, while $z_i(\mu)=0$ for
$i=3-10$. From $\mu=m_c$ down, as the charm-induced penguins come into play,
$z_i(\mu)$ evolve, given the proper matching conditions,
with the full anomalous dimension matrix.
The Wilson coefficients $v_i(\mu)$ ($y_i (\mu) = v_i (\mu) - z_i (\mu)$)
arise at $m_W$ due to integration of the $W$ and top quark fields.
They coincide with $z_i(\mu)$ for $i=1,2$,
the information about the
top quark being encoded in the $i=3-10$ components.
Finally, $\lambda_i \equiv
V_{id}V^*_{is}$, where $V$ is the Kobayashi-Maskawa (KM) matrix, and
$\tau \equiv  - \lambda_t/\lambda_u$.

Let us mention that, according to  standard conventions,
Im $\lambda_u =0$ and the short distance component of $\epsilon'/\epsilon$
is thus determined by the Wilson coefficients $y_i$. Notice that,
following the approach of
ref.~\cite{Buras}, $y_1(\mu)=$ $y_2(\mu)=0$ and the effect of $Q_{1,2}$ appears
only through the linearly dependent operators $Q_{4,9,10}$.

In this letter,
we would like to discuss the possible relevance of two extra
operators that have  so far been  neglected, namely
\beq
Q_{11} = \frac{g_s}{16\pi^2} \, m_s \: \bar{s}\, \sigma_{\mu\nu}
t^a G^{\mu\nu}_a (1-\gamma_5 )\, d  \label{Q11}
\eeq
and
\beq
Q_{12} = \frac{eQ_d}{16\pi^2}\, m_s \: \bar{s}\, \sigma_{\mu\nu}
 F^{\mu\nu} (1-\gamma_5 )\, d \label{Q12} \, ,
\eeq
which account for the magnetic and electric dipole part of,
respectively, the QCD and electromagnetic penguin operators.
In \eq{Q11}, $t^a$ are the $SU(3)_c$ generators normalized
as $\Tr (t^at^b) = (1/2)\delta^{ab}$, while in \eq{Q12} $Q_d = -1/3$
is the charge of the down quarks; $\sigma_{\mu\nu} = (i/2) [\gamma_\mu , \,
\gamma_\nu]$.

These operators have been left out in the past, together with
$Z$ penguins and electroweak box diagrams, because their
Wilson coefficients exhibit, for light quark masses, a power-like
GIM suppression \cite{GIM}, to be compared with the leading soft
logarithmic behavior of the ``monopole'' component of the
gluonic and photonic penguins. However, for a heavy top quark, there is no
reason a priori to discard these contributions, which should be
considered on the same grounds as
$Z$ penguins and electroweak box diagrams. In addition, $Q_{11}$
and $Q_{12}$
receive a large QCD renormalization due to the mixings with the other
operators, as has been shown in the case of
$b\to s\gamma$ \cite{qcdbsgamma}.

The potential relevance of the
dipole components of the gluonic and photonic penguins for
$\epsilon'/\epsilon$
has been emphasized by one of us a few years ago~\cite{Stefano}.
Only recently has a quantitative attempt to estimate the relevance
of $Q_{11}$  appeared~\cite{russi}. However, in ref.~\cite{russi}, only
the multiplicative renormalization---which is by no means the leading
effect---is  taken into account. We also  disagree
on the evaluation of the   hadronic matrix element that, we believe, is
overestimated in ref.~\cite{russi}  by more than one order of magnitude (by
a factor 8 using their method).

The evaluation of the matrix elements is in fact the crucial point
in determining the relevance of the two additional operators.
As we shall see, while the contribution of the electromagnetic
penguin $Q_{12}$ seems to be vanishingly small, that of
the gluonic  $Q_{11}$ should not be neglected. In fact,
its Wilson coefficient turns out to be sizeably enhanced by mixing
and the matrix element is not dramatically
suppressed with respect to the other ten operators.

\bigskip

{\bf 2.} In order to compare meaningfully the relevance of the
new contributions with the traditional ones,
from now on we  strictly follow the analysis
presented in ref.~\cite{Buras}. The parameter $\epsilon'/\epsilon$ can be
obtained by means of the Hamiltonian in \eq{ham}, which yields \cite{Buras}
\beq
\frac{\epsilon '}{\epsilon} = 10^{-4} \left[ \frac{ \mbox{Im}\, \lambda
_t}{1.7 \times 10^{-4}} \right] \Bigl[ P^{(1/2)} - P^{(3/2)} \Bigr] \, ,
\label{eps}
 \eeq
where
\bea
 P^{(1/2)} & = & r\ \sum y_i \, \langle 2 \pi,\ I=0 |\, Q_i \, |  K^0 \rangle
\left(1 - \Omega_{\eta +\eta'} \right) \\
 P^{(3/2)} & = & \frac{r}{\omega} \sum y_i \, \langle 2 \pi,\ I=2 |\,  Q_i
\, | K^0 \rangle  \ .
\eea
We take, as input values for
the relevant quantities, the central values given in  appendix C of
ref. \cite{Buras}. This will allow us to reproduce in the ten-operator case
the central values of the
leading order (LO) results, as given in appendix B of ref.~\cite{Buras}.
In particular, we take
\beq
r = 1.7\frac{G_F\omega}{2\mod{\epsilon}\Re{A_0}} \simeq 594 \ \mbox{GeV}^{-3},
\qquad \omega = 1/22.2\ , \qquad \Omega_{\eta + \eta'} = 0.25 \ ;
\eeq
$\Im{\lambda_t}$ is determined from the experimental value of
$\epsilon$ as an interpolating function of $m_t$.
For instance, taking
the KM phase $\delta_{KM}$ in the first quadrant, we find,
for the central value,
\beq
\Im{\lambda_t} = 2.77\times 10^{-4}\ x_t^{-0.365}\, ,
\label{Imlamti1}
\eeq
where $x_t = m_t^2/m_W^2$.

 The value of the Wilson coefficients $y_{11}$ and $y_{12}$ at
the hadronic scale of 1 GeV can be found by means of the renormalization
group equations. Denoting generically by $\vec C(\mu)$ the vector of
Wilson coefficients, its scale dependence is given by
\beq
\left[\mu\frac{\partial}{\partial\mu}
+ \beta(g)\frac{\partial}{\partial g}\right]
\vec C \left( \frac{m_W^2}{\mu^2},g^2,\alpha \right) \ \ =\ \
\hat\gamma^T(g^2,\alpha)\ \vec C \left( \frac{m_W^2}{\mu^2},g^2,\alpha
\right) \, ,
\label{RGE}
\eeq
where $\beta(g)$ is the QCD beta function and $\alpha$ the electromagnetic
coupling (the running of $\alpha$ is being neglected).
At the leading order we have
\beq
\hat\gamma(g^2,\alpha) = \frac{\alpha_s}{4\pi}\hat\gamma_s^{(0)}
                       + \frac{\alpha}{4\pi}\hat\gamma_e^{(0)} \, ,
\label{gammas}
\eeq
where $\hat\gamma_s^{(0)}$ governs the QCD and $\hat\gamma_e^{(0)}$ the
electromagnetic running.

The $10\times 10$ mixing matrix of the anomalous dimension for
 $Q_{1-10}$
has recently been  evaluated at the next-to-leading order in
refs.~\cite{Buras,Roma}.   On the other hand, the matrix of the strong
anomalous
dimensions of $Q_{11}$ and $Q_{12}$, and their QCD-induced mixing with
$Q_{1-6}$, can be borrowed from the existing calculations for the $b\to
s\gamma$
decay \cite{qcdbsgamma} (we will use the most recent and complete analysis of
ref.~\cite{matrice}).  Assembling the known results, we thus
 write the leading order
$12\times 12$ anomalous dimension matrix $\hat\gamma_s^{(0)\ T}$ as
\begin{minipage}[t]{15.2cm} {
\scriptsize
$$
 \left(
\matrix{ -6/N & 6 & 0 & 0 & 0 \cr
6 & -6/N & 0 & 0 & 0 \cr
0 & -2/3 N & -22/3 N & 6 - 2 n_f/3 N & 0 \cr
0 & 2/3 & 22/3 & -6/N + 2 n_f/3 & 0  \cr
0 & -2/3 N & -4/3 N &  - 2 n_f/3 N & 6/N \cr
0 & 2/3 & 4/3 & 2 n_f/3 &  -6  & \cr
 0 & 0 & 0 & 0 & 0  \cr
 0 & 0 & 0 & 0 & 0 \cr
 0 & 0 & 0 & 0 & 0  \cr
 0 & 0 & 0 & 0 & 0  \cr
 6 & 22 N/9 -58/9N & 44N/9 - 116/9N +6n_f & 12 + 22 N n_f/9 -
58 n_f/9N & -4N +8/N - 6n_f  \cr
 0 & (12 Q_u/Q_d +8/9)C_f & 232 C_f/9 & (12 \bar{n}_f + 8 n_f/9 )C_f &
-16 C_f   \cr }
\right.
$$ }
\beq
\hfill
\left.
{\scriptsize
\matrix{0 & 0 & 0 & 0 & 0 & 0 & 0 \cr
 0 & 0 & 0 & 0 & 0 & 0 & 0 \cr
-2 n_f/3 N & 0 & \bar{n}_f/3 N & 2/3 N & \bar{n}_f/3 N & 0 & 0 \cr
 2 n_f/3  & 0 & -\bar{n}_f/3 & -2/3  & -\bar{n}_f/3  & 0 & 0 \cr
-2 n_f/3 N & 0 & \bar{n}_f/3 N & 2/3 N & \bar{n}_f/3 N & 0 & 0 \cr
-12C_f + 2n_f/3 & 0 & -\bar{n}_f/3  & -2/3 & -\bar{n}_f/3 & 0 & 0 \cr
 0 & 6/N & 0 & 0 & 0 & 0 & 0 \cr
 0 & -6 & -12C_f & 0 & 0 & 0 & 0 \cr
 0 & 0 & 0 & -6/N & 6 & 0 & 0 \cr
 0 & 0 & 0 & 6 & -6/N & 0 & 0 \cr
-8 -32N n_f/9 + 50 n_f/9N &  X_7 & X_8 & X_9 & X_{10} & 4N - 8/N & 0 \cr
(-12 \bar{n}_f + 8 n_f/9 ) C_f & Y_7 & Y_8 & Y_9 & Y_{10} & 8C_f & 8C_f \cr}
}
\right)
\label{matricione}
\eeq
\bigskip
\end{minipage}
where $N$ is the number of colors,
$n_f$ is the number of active flavors,
and $\bar{n}_f = n_d + (Q_u/Q_d) n_u$,
with $Q_u = 2/3$, $Q_d = -1/3$, $n_u$ ($n_d$) being
the number of up (down) active quarks; finally, $C_f \equiv
(N^2-1)/2N$.

A few comments are in order. The $10 \times 10$ part of the matrix
(\ref{matricione}) is identical to that used in refs.~\cite{Buras,Roma}
for the leading one-loop order result.
The two extra columns and rows represent the mixing of the
first ten operators with the two new ones, which takes place first at the
two-loop level. We have taken for these entries
the 't Hooft-Veltman (HV) scheme results~\cite{Buras,Roma}. In this way,
no finite additional contributions to the renormalization of
$y_{11}$ and $y_{12}$ arise at the various quark thresholds (for a discussion
see ref.~\cite{matrice}).
The entries labelled with $X_{7-10}$, $Y_{7-10}$ represent
the mixings of, respectively,
$Q_{11}$ and $Q_{12}$ with the electroweak penguins,
which have not yet been computed. We set them equal to zero, expecting
that the leading contribution to the running of the Wilson coefficients
$y_{11}$ and $y_{12}$ arises from the mixing with the current-current
operators and the gluonic penguins.

Since we want to fully reproduce the leading-order results for the $10\times
10$
operator system, we have included, in the running of the Wilson
coefficients, the effect of the electromagnetic renormalization
following the procedure described in ref.~\cite{Buras}. To
the  $10\times 10$ electromagnetic
anomalous dimension matrix, which can be found in refs.~\cite{Buras,Roma},
we have simply
added two rows and two columns of zeros, thus
neglecting the electromagnetic running of $Q_{11}$ and $Q_{12}$.

The fact that the mixing of $Q_{11}$ and $Q_{12}$ with the other operators
arises at the two-loop level introduces an explicit scheme dependence
in our analysis. From the next-to-leading results of refs.~\cite{Buras,Roma},
it appears however that the inclusion of next-to-leading effects reduces
the values of $\epsilon'/\epsilon$ compared to the leading-order result.
Since the presence of the two extra  operators does not affect the
running of $y_{1-10}(\mu)$ (see \eqs{RGE}{matricione}),
we conclude that our simplified leading-order
procedure will at most
underestimate the effective weight in $\epsilon'/\epsilon$
of the new contributions.

For the purpose of comparison, we take
as initial conditions for $v_{1-10}(m_W)$ the leading-order values
of ref.~\cite{Buras}, which coincide with the HV results when
neglecting $O(\alpha_s)$ corrections to $v_{1,2}(m_W)$
(strictly speaking at the leading order one should set $v_{3-10}(m_W)=0$).
For what concerns the new coefficients $v_{11,12}(m_W)$ we have
\beq
v_{11}(m_W) =
-E'(m_t^2/m_W^2) \qquad \mbox{and} \qquad v_{12}(m_W) = -D'(m_t^2/m_W^2)/Q_d
\, ,
\eeq
where
\bea
E'(x) & = & \frac{3x^2 }{2 (1 - x)^4} \ln x
-\frac{x^3 - 5 x^2 -2 x}{4 (1 - x)^3} \\
D'(x) & = & \frac{x^2 (2 - 3 x)}{2 (1 - x)^4} \ln x
-\frac{8 x^3 + 5 x^2 -7 x}{12 (1 - x)^3} \, .
\eea

In Table 1 we report the results for $z_{1-12}$(1 GeV)
and $y_{3-12}$(1 GeV) (remember that $y_{1,2}(\mu)=0$)
compared to their initial values, for $m_t$ equal to 130 and 170 GeV and
$\Lambda_{QCD}^{(4)}$ equal to 200 and 300 MeV.
As expected,
a comparison between our results and those of Table 1 of ref.~\cite{Buras}
shows that we fully agree on the values of the renormalized coefficients
for the first ten operators.
Regarding $v_{11,12}$ ($= y_{11,12} + z_{11,12}$),
we note that the effect of operator mixing induces a
renormalization that is a factor of 4--5 larger than that
induced by multiplicative running alone (which roughly reduces
by a factor of 2 the initial Wilson coefficients).

In order to ascertain the importance of $Q_{11}$ and $Q_{12}$
in the estimate of
$\epsilon'/\epsilon$, we now have  to address the
question of the evaluation of their hadronic matrix elements.
\begin{table}
{\footnotesize
\begin{tabular}{|c|c c|c c||c c|c c|}
\hline
$\Lambda_{QCD}^{(4)}$ & \multicolumn{4}{c||}{ $0.2$ GeV }
                      & \multicolumn{4}{c|}{ $0.3$ GeV } \\
\hline
$m_t$&\multicolumn{2}{c|}{$130$ GeV}&\multicolumn{2}{c||}{$170$ GeV}
     &\multicolumn{2}{c|}{$130$ GeV}&\multicolumn{2}{c|}{$170$ GeV}  \\
\hline \hline
$z_1$&\multicolumn{4}{c||}{$-0.587\ (0.0)$}
     &\multicolumn{4}{c|}{$-0.715\ (0.0)$} \\
\hline
$z_2$&\multicolumn{4}{c||}{$1.319\ (1.0)$}
     &\multicolumn{4}{c|}{$1.409\ (1.0)$} \\
\hline\hline
$z_3$&\multicolumn{4}{c||}{$0.004$}&\multicolumn{4}{c|}{$0.005$} \\
\hline
$z_4$&\multicolumn{4}{c||}{$-0.010$}&\multicolumn{4}{c|}{$-0.014$} \\
\hline
$z_5$&\multicolumn{4}{c||}{$0.003$}&\multicolumn{4}{c|}{$0.004$} \\
\hline
$z_6$&\multicolumn{4}{c||}{$-0.011$}&\multicolumn{4}{c|}{$-0.015$} \\
\hline\hline
$z_7/\alpha$&\multicolumn{4}{c||}{$0.005$}&\multicolumn{4}{c|}{$0.008$} \\
\hline
$z_8/\alpha$&\multicolumn{4}{c||}{$0.001$}&\multicolumn{4}{c|}{$0.001$} \\
\hline
$z_9/\alpha$&\multicolumn{4}{c||}{$0.005$}&\multicolumn{4}{c|}{$0.009$} \\
\hline
$z_{10}/\alpha$&\multicolumn{4}{c||}{$-0.001$}&\multicolumn{4}{c|}{$-0.001$} \\
\hline \hline
$z_{11}$&\multicolumn{4}{c||}{$-0.038$}&\multicolumn{4}{c|}{$-0.045$} \\
\hline
$z_{12}$&\multicolumn{4}{c||}{$0.431$}&\multicolumn{4}{c|}{$0.582$} \\
\hline
\hline
$y_3$&$0.027$&$(-0.0003)$&$0.028$&$(0.0004)$&$0.034$&$(-0.0003)$
&$0.035$&$(0.0004)$ \\
\hline
$y_4$&$-0.048$&$(0.0018)$&$-0.049$&$(0.0014)$&$-0.056$&$(0.0020)$
&$-0.057$&$(0.0015)$ \\
\hline
$y_5$&$0.011$&$(-0.0006)$&$0.012$&$(-0.0005)$&$0.012$&$(-0.0007)$
&$0.013$&$(-0.0005)$ \\
\hline
$y_6$&$-0.078$&$(0.0018)$&$-0.080$&$(0.0014)$&$-0.102$&$(0.0020)$
&$-0.104$&$(0.0015)$ \\
\hline\hline
$y_7/\alpha$&$-0.025$&$(0.091)$&$0.025$&$(0.151)$&$-0.017$&$(0.091)$
&$0.033$&$(0.151)$ \\
\hline
$y_8/\alpha$&$0.053$&$(0.0)$&$0.109$&$(0.0)$&$0.074$&$(0.0)$&$0.148$&$(0.0)$ \\
\hline
$y_9/\alpha$&$-1.160$&$(-0.793)$&$-1.554$&$(-1.094)$&$-1.222$&$(-0.793)$
&$-1.644$&$(-1.094)$ \\
\hline
$y_{10}/\alpha$&$0.488$&$(0.0)$&$0.663$&$(0.0)$&$0.592$&$(0.0)$&$0.806$
&$(0.0)$ \\
\hline \hline
$y_{11}$&$-0.328$&$(-0.172)$&$-0.340$&$(-0.193)$&$-0.350$&$(-0.172)$
&$-0.362$&$(-0.193)$ \\
\hline
$y_{12}$&$2.070$&$(0.965)$&$2.161$&$(1.158)$&$2.166$&$(0.965)$
&$2.245$&$(1.158)$ \\
\hline
\end{tabular}
}
\caption{Wilson coefficients at $\mu=1$ GeV, in the leading order,
as explained in the text ($\alpha=1/128$).
The corresponding values at $\mu=m_W$ are given in parenthesis.
In addition, at $\mu=m_c$ we have $z_{3-12}(m_c)=0$.  }
\end{table}

\newpage

{\bf 3.}
Most of the  uncertainties in the estimate of
$\epsilon'/\epsilon$ arise from the evaluation
of the hadronic matrix elements.
For the operators $Q_{1-10}$, we follow the
strategy of ref.~\cite{Buras}
where the various matrix elements are evaluated by means of
the $1/N$ expansion and soft-meson methods. Overall
coefficients $B_i^{(1/2)}$ and $B_i^{(3/2)}\ (i = 1-10)$
parameterize our level of ignorance of their normalization scale, scale
dependence and error estimate. Since some of the matrix elements
are phenomenologically determined using $CP$-conserving data
($\Delta I=1/2$ rule),
and further relations among the $B_i$'s
are advocated, the final result
is parameterized in terms of two coefficients:
$B_6^{(1/2)}$ and $B_8^{(3/2)}$, which take 1 as central value.
The inclusion of $Q_{11}$ and $Q_{12}$ requires three additional effective
parameters: $B_{11}^{(1/2)}$, $B_{12}^{(1/2)}$ and $B_{12}^{(3/2)}$.

In order to evaluate $\langle \pi^+ \pi^- | \, Q_{11} \, | K^0 \rangle$ and
$\langle \pi^+ \pi^- | \, Q_{12} \, | K^0
\rangle$, we use effective chiral lagrangian techniques. The lowest order
 chiral representation for these operators vanishes because of the cancellation
between the direct transition and the pole diagram induced by a
non-vanishing $\langle 0 | \, Q_{11} \, | K^0 \rangle$ matrix
element~\cite{DH}. The leading
next order ($O\,(p^4)$) contribution arises from the bosonization of
$Q_{11}$ into
$L_{Q_{11}}$, where
 \beq
L_{Q_{11}} = G_{Q_{11}} \: \Tr \left\{ \left[ \Sigma^{\dag} {\cal M}_q
\lambda_+ +
 \lambda_+ \Sigma {\cal M}_q^{\dag} \right] \partial _\mu \Sigma^{\dag}
  \partial^\mu \Sigma \right\} \ . \label{lag}
\eeq
In the formula above $\lambda_+ =( \lambda_6 + i \lambda_7)/2$
is the  octet $\Delta S=1$ projector,
${\cal M}_q$ is the quark current mass matrix, $\Sigma = \exp \left(
\sqrt{2}\, i
\,\Pi/F_\pi \right)$ and $\Pi = \sum_{a=1}^8 \lambda^a \pi^a$ is the usual
matrix of chiral SU(3) Goldstone bosons.

The constant $G_{Q_{11}}$ can be computed by considering a model for the QCD
effective action at long distances~\cite{model}; it is found to be~\cite{Eeg}
\beq
G_{Q_{11}} = - \frac{11}{64\pi^2} \langle 0 |  \bar{q} q | 0 \rangle \, ,
\eeq
where $ \langle 0 |  \bar{q} q | 0 \rangle$ is the quark condensate that we
take to
be $-F_K^2 m_K^2/[2(m_s+m_u)]$. More details on this computation will be given
elsewhere~\cite{BEF}. The final result can be written as
\beq
\langle \pi^+ \pi^- | \, Q_{11} \, | K^0 \rangle =
- \frac{1}{16 \pi^2} \frac{11}{2} \frac{ m_s}{m_s +m_u}\:
\frac{F_K^2}{F_{\pi}^3}\, m_K^2 \, m_\pi^2\: B_{11}^{(1/2)} \, , \label{mat}
\eeq
with $B_{11}^{(1/2)} =1$.

The corresponding matrix element for $Q_{12}$ is very much suppressed because
it is proportional to the photon condensate and it must therefore be very
small, if
not vanishing; we will neglect it altogether in what follows.
Nevertheless,  we must bear in mind the possibility that the same
electromagnetic dipole penguin operator
could give a sizeable contribution, for instance,
once saturated with an external quark current and written as a four-fermion
operator.  Such a possibility is under investigation.

Notice in eq.~(\ref{mat}) the presence
 of the factor $1/16\pi^2$, reminding  us
that the leading-order mixing between $Q_{1-10}$ and $Q_{11}$ appears
at the two-loop level.

Since $\vev{\pi^+\pi^-|Q_{11}| K^0} = \vev{\pi^0\pi^0|Q_{11}| K^0}$
we finally obtain
\bea
\langle 2 \pi , \: I=0 |\, Q_{11} \, | K^0 \rangle  & = &
\sqrt{\frac{3}{2}} \: \langle \pi^+ \pi^-
| Q_{11} | K^0 \rangle \nnu \\
\langle 2\pi , \: I=2 |\,  Q_{11}\, | K^0 \rangle  & = & 0 \, ,
\eea
which enter in the determination of $\epsilon '/\epsilon$ (see
eq.~(\ref{eps})).

We feel that our estimates for the hadronic matrix elements of the operator
$Q_{11}$ and $Q_{12}$ are  consistent with those in
ref.~\cite{Buras} for the other ten
operators.

\bigskip

{\bf 4.} We can now discuss the effect of the various operators in
determining the size of $\epsilon'/\epsilon$, and
identify the role of the new contributions.
We follow here the analysis of ref.~\cite{Buras}, where the
hadronic matrix elements are evaluated using the $1/N$ expansion
and soft-meson methods. As previously mentioned, the final results
are written in terms of effective coefficients $B_i^{(1/2)}$ and
$B_i^{(3/2)}$, which encompass our lack of knowledge on scale
normalization, scale dependence and goodness of the method.
The matrix elements of $Q_1$ and $Q_2$ can however be determined
phenomenologically from the experimental values of $\Re{A_0}$
and $\Re{A_2}$, so as to reproduce the $\Delta I= 1/2$ rule.
In particular, in ref.~\cite{Buras} it is found that
$B_{2,LO}^{(1/2)}\approx 5.8$, which is about three times larger
that the $1/N$ result. Related to this coefficient is the value
of $B_{1,LO}^{(1/2)}$, which we find equal to 19.3 and 13.5 for
$\Lambda_{QCD}^{(4)}=0.2$ and 0.3 GeV, respectively. Correspondingly,
$B_{1,LO}^{(3/2)}=0.48$ and 0.50.
Relations among the other coefficients are advocated in ref.~\cite{Buras},
depending on the relevance and the role of the various
operators, so as to reduce, in the ten-operator case, the description
of the hadronic sector to two effective parameters: $B_{6}^{(1/2)}$ and
$B_{8}^{(3/2)}$, whose $1/N$ value is 1.

Since the determination of
$B_{1}$ and $B_{2}$ is best achieved at $\mu=m_c$ \cite{Buras},
all
the hadronic matrix elements are assumed to be evaluated at that scale
and renormalized down to 1 GeV via their anomalous dimension matrix.

We proceed analogously by setting $B_{11}^{(1/2)}=1$ and
$B_{12}^{(1/2)}=B_{12}^{(3/2)}=0$ at $\mu=m_c$ and using the
$12\times 12$ QCD and electromagnetic evolution matrices
to evolve all the hadronic matrix elements to the 1 GeV scale.
Since the anomalous dimension matrices, which govern the
evolution of the hadronic matrix elements, are the transpose
of those evolving the Wilson coefficients, we now find  that
the presence of $Q_{11,12}$ affects the renormalization
of the first ten operators. On the other hand, the evolution
of $Q_{11,12}$ is determined solely by their $2\times 2$
anomalous dimension matrix, which imply that
the matrix element of $Q_{12}$ remains vanishing.

As a consequence of the previous remarks,
our results for the individual contributions of the operators $Q_{1-10}$
to $\epsilon'/\epsilon$ differ (slightly) from those
reported in ref.~\cite{Buras}.
We have however checked that, in the ten-operator case, we reproduce
 their leading-order results exactly.

We have chosen to illustrate numerically our conclusions
as tables.
Tables 2, 3 and 4 show the contributions to $\epsilon '/\epsilon$
of each operator, for different choices of $\Lambda_{QCD}$, $B_6^{(1/2)}$
and $B_8^{(3/2)}$.
The first ten contributions are also partially grouped in a ``positive''
gluonic component versus a ``negative'' electroweak component, which
shows the ``superweak'' behavior of $\epsilon'/\epsilon$ within the standard
model, as
the top mass increases.
The total effect in the twelve-operator case is then compared
with the ten-operator case result (we have filled in those
data that are not explicitly available from ref.~\cite{Buras}).

These tables suggests that
$Q_{11}$ should be consistently included in any estimate of
$\epsilon'/\epsilon$. Its sign is the same as that of $Q_6$ and therefore
makes $\epsilon'/\epsilon$ larger;
for large $m_t$,  the ``super-weak'' regime
is accordingly shifted to higher values.

Let us conclude by mentioning the importance of
completing the leading-order
(two-loop) calculation of the anomalous dimension matrices---which is important
also for the $\Delta B =1$ processes.
Moreover,
a  consistent discussion of the short-distance
part of the present analysis should include the next-to-leading order.
This however implies computing the mixings of $Q_{11,12}$ with
the other operators at the three-loop level---a truly formidable task.
\bigskip

We thank J.O. Eeg, E. Franco, S. Narison, P. Nason, N. Paver, A. Pich
and Riazuddin for helpful discussions.
\clearpage
\begin{table}
\begin{tabular}{|c||c|c||c|c||c|c||c|c|}
\hline
\multicolumn{9}{|c|}{$\epsilon'/\epsilon\times 10^{4}$\ \ (Leading Order)} \\
\hline
$m_t$ &\multicolumn{2}{c||}{ $130$ GeV} & \multicolumn{2}{c|}{ $170$ GeV}
      &\multicolumn{2}{c||}{ $200$ GeV} & \multicolumn{2}{c|}{ $230$ GeV}
 \\ \hline \hline
$Q_3$ & $0.5$ &  & $0.4$ &  & $0.4$ &  & $0.3$ & \\
\cline{1-2} \cline{4-4} \cline{6-6} \cline{8-8}
$Q_4$ & $-5.2$ &  & $-4.4$ &  & $-3.9$&  & $-3.6$ & \\
\cline{1-2}  \cline{4-4}  \cline{6-6}  \cline{8-8}
$Q_5$ & $-0.6$ &9.4 & $-0.5$ & 7.8 & $-0.4$ & 7.2 & $-0.4$ & 6.4 \\
\cline{1-2} \cline{4-4} \cline{6-6} \cline{8-8}
$Q_6$ & $14.7$ &  & $12.3$ &  & $11.0$ &  & $10.1$ & \\
\hline \hline
$Q_7$ & $0.2$ &  & $-0.4$ &  & $-0.7$ &  & $-1.1$ & \\
\cline{1-2} \cline{4-4} \cline{6-6} \cline{8-8}
$Q_8$ & $-3.6$ &  & $-5.9$ &  & $-7.6$&  & $-9.2$ & \\
\cline{1-2}  \cline{4-4} \cline{6-6} \cline{8-8}
$Q_9$ & $3.0$ & $-1.5$  & $3.4$ & $-4.1$  & $ 3.6$ & $-6.0$ & $3.8$ &$-7.9$ \\
\cline{1-2} \cline{4-4} \cline{6-6} \cline{8-8}
$Q_{10}$ & $-1.1$ &  & $-1.2$ &  & $-1.3$ &  & $-1.4$ & \\
\hline \hline
$Q_{11}$ & \multicolumn{2}{r||}{$0.5$} & \multicolumn{2}{r||}{$0.4$} &
\multicolumn{2}{r||}{$0.3$} & \multicolumn{2}{r|}{$0.3$} \\
\hline \hline
$Q_1 -Q_{12}$ & \multicolumn{2}{r||}{$8.4$} & \multicolumn{2}{r||}{$4.2$} &
\multicolumn{2}{r||}{$1.4$} & \multicolumn{2}{r|}{$-1.2$}
\\ \hline
Ref.~\cite{Buras} & \multicolumn{2}{r||}{$8.0$} & \multicolumn{2}{r||}{$3.8$}
                  & \multicolumn{2}{r||}{$1.0$} & \multicolumn{2}{r|}{$-1.5$}
\\ \hline
\end{tabular}
\caption{ Anatomy of $\epsilon'/\epsilon$ for
$\Lambda_{QCD}^{(4)} = 300$ MeV,
$B^{(1/2)}_6 =  B^{(3/2)}_8 = B^{(1/2)}_{11} =1$.
The contribution of each operator is shown at $\mu=1\ GeV$,
together with partial
grouping of the gluonic and electroweak sectors.
The contribution of $Q_{12}$ is being neglected.
The total effect
is compared with the corresponding leading-order result of the ten-operator
case
(last line).}
\end{table}
\clearpage
\begin{table}
\begin{tabular}{|c||c|c||c|c||c|c||c|c|}
\hline
\multicolumn{9}{|c|}{$\epsilon'/\epsilon\times 10^{4}$\ \ (Leading Order) } \\
\hline
$m_t$ &\multicolumn{2}{c||}{ $130$ GeV} & \multicolumn{2}{c|}{ $170$ GeV}
      &\multicolumn{2}{c||}{ $200$ GeV} & \multicolumn{2}{c|}{ $230$ GeV}
 \\ \hline \hline
$Q_3$ & $0.4$ &  & $0.3$ &  & $0.3$ &  & $0.3$ & \\
\cline{1-2} \cline{4-4} \cline{6-6} \cline{8-8}
$Q_4$ & $-5.1$ &  & $-4.2$ &  & $-3.8$&  & $-3.5$ & \\
\cline{1-2}  \cline{4-4}  \cline{6-6}  \cline{8-8}
$Q_5$ & $-0.6$ &6.7 & $-0.5$ & 5.6 & $-0.4$ & 5.1 & $-0.4$ &4.6 \\
\cline{1-2} \cline{4-4} \cline{6-6} \cline{8-8}
$Q_6$ & $12.0$ &  & $10.0$ &  & $9.0$ &  & $8.2$ & \\
\hline \hline
$Q_7$ & $0.4$ &  & $-0.3$ &  & $-0.7$ &  & $-1.2$ & \\
\cline{1-2} \cline{4-4} \cline{6-6} \cline{8-8}
$Q_8$ & $-2.7$ &  & $-4.6$ &  & $-5.9$&  & $-7.3$ & \\
\cline{1-2}  \cline{4-4} \cline{6-6} \cline{8-8}
$Q_9$ & $2.9$ & $-0.3$  & $3.2$ & $-2.7$  & $ 3.5$ & $-4.2$ & $3.7$ &$-5.9$ \\
\cline{1-2} \cline{4-4} \cline{6-6} \cline{8-8}
$Q_{10}$ & $-0.9$ &  & $-1.0$ &  & $-1.1$ &  & $-1.1$ & \\
\hline \hline
$Q_{11}$ & \multicolumn{2}{r||}{$0.4$} & \multicolumn{2}{r||}{$0.4$} &
\multicolumn{2}{r||}{$0.3$} & \multicolumn{2}{r|}{$0.3$} \\
\hline \hline
$Q_1 -Q_{12}$ & \multicolumn{2}{r||}{$6.8$} & \multicolumn{2}{r||}{$3.3$} &
\multicolumn{2}{r||}{$1.0$} & \multicolumn{2}{r|}{$-1.1$}
\\ \hline
Ref.~\cite{Buras} & \multicolumn{2}{r||}{$6.4$} & \multicolumn{2}{r||}{$3.0$}
                  & \multicolumn{2}{r||}{$0.7$} & \multicolumn{2}{r|}{$-1.4$}
\\ \hline
\end{tabular}
\caption{Same as in Table 2 for $\Lambda_{QCD}^{(4)} = 200$ MeV,
$B^{(1/2)}_6 = B^{(3/2)}_8 =1$.}
\end{table}
\clearpage
\begin{table}
\begin{tabular}{|c||c|c||c|c||c|c||c|c|}
\hline
\multicolumn{9}{|c|}{$\epsilon'/\epsilon\times 10^{4}$\ \ (Leading Order) } \\
\hline
$m_t$ &\multicolumn{2}{c||}{ $130$ GeV} & \multicolumn{2}{c|}{ $170$ GeV}
      &\multicolumn{2}{c||}{ $200$ GeV} & \multicolumn{2}{c|}{ $230$ GeV}
 \\ \hline \hline
$Q_3$ & $0.5$ &  & $0.4$ &  & $0.4$ &  & $0.4$ & \\
\cline{1-2} \cline{4-4} \cline{6-6} \cline{8-8}
$Q_4$ & $-5.2$ &  & $-4.4$ &  & $-4.0$&  & $-3.6$ & \\
\cline{1-2}  \cline{4-4}  \cline{6-6}  \cline{8-8}
$Q_5$ & $-0.4$ &5.8 & $-0.4$ & 4.8 & $-0.3$ & 4.1 & $-0.3$ & 3.9 \\
\cline{1-2} \cline{4-4} \cline{6-6} \cline{8-8}
$Q_6$ & $11.0$ &  & $9.2$ &  & $8.2$ &  & $7.5$ & \\
\hline \hline
$Q_7$ & $0.2$ &  & $-0.3$ &  & $-0.6$ &  & $-0.8$ & \\
\cline{1-2} \cline{4-4} \cline{6-6} \cline{8-8}
$Q_8$ & $-2.6$ &  & $-4.3$ &  & $-5.6$&  & $-6.8$ & \\
\cline{1-2}  \cline{4-4} \cline{6-6} \cline{8-8}
$Q_9$ & $3.0$ & $-0.5$  & $3.4$ & $-2.4$  & $ 3.6$ & $-3.6$ & $3.9$ &$-5.1$\\
\cline{1-2} \cline{4-4} \cline{6-6} \cline{8-8}
$Q_{10}$ & $-1.1$ &  & $-1.2$ &  & $-1.3$ &  & $-1.4$ & \\
\hline \hline
$Q_{11}$ & \multicolumn{2}{r||}{$0.5$} & \multicolumn{2}{r||}{$0.4$} &
\multicolumn{2}{r||}{$0.3$} & \multicolumn{2}{r|}{$0.3$} \\
\hline \hline
$Q_1 -Q_{12}$ & \multicolumn{2}{r||}{$5.7$} & \multicolumn{2}{r||}{$2.8$} &
\multicolumn{2}{r||}{$0.8$} & \multicolumn{2}{r|}{$-1.0$}
\\ \hline
Ref.~\cite{Buras} & \multicolumn{2}{r||}{$5.3$} & \multicolumn{2}{r||}{$2.4$}
                  & \multicolumn{2}{r||}{$0.5$} & \multicolumn{2}{r|}{$-1.3$}
\\ \hline
\end{tabular}
\caption{Same as in Table 2 for $\Lambda_{QCD}^{(4)} = 300$ MeV,
$B^{(1/2)}_6 = B^{(3/2)}_8 =0.75$.}
\end{table}

\clearpage
\renewcommand{\baselinestretch}{1}


\begin{thebibliography}{99}
{\small


\bibitem{epsilon} A.J. Buras and M.K. Harlander,
 in {\em Heavy Flavours},
eds. A.J. Buras and M. Lindner (World Scientific, Singapore, 1992).

\bibitem{Lisa} M. Lusignoli, \npb{325}{89}{33};\\
J.M. Flynn and L. Randall, \plb{224}{89}{221};\\
G. Buchalla, A.J. Buras and M.K. Harlander, \npb{337}{90}{313};\\
E.A. Paschos and Y.L. Wu, \mpla{6}{91}{93};\\
M. Lusignoli \etal, \npb{369}{92}{139}.

\bibitem{Buras} A.J. Buras, M. Jamin and M.E. Lautenbacher, \npb{408}{93}{209};
\\ see also: G. Buchalla \etal, in ref.~\cite{Lisa}.

\bibitem{Roma} M. Ciuchini, E. Franco, G. Martinelli and L. Reina,
preprint LPTENS 93/11 (April 1993);
\plb{301}{93}{263}.

\bibitem{experiments} G.D. Barr \etal ~(NA31), \plb{317}{93}{233};\\
L.K. Gibbons \etal ~(E731),
\prl{70}{93}1203.

\bibitem{GIM} S.L Glashow, J. Iliopoulos and L. Maiani, \prd{2}{70}{1285}.

\bibitem{qcdbsgamma}
S. Bertolini, F. Borzumati and A. Masiero, \prl{59}{87}{180}; \\
N.G. Deshpande, P. Lo, J. Trampetic,
  G. Eilam and P. Singer, \prl{59}{87}{183};       \\
B. Grinstein, R. Springer and M. Wise,
  \plb{202}{88}{138}; \npb{339}{90}{269};      \\
R. Grigjanis, P.J. O'Donnell, M. Sutherland and H. Navelet,
\plb{213}{88}{355}, (E) \ibid{286}{92}{413};           \\
G. Cella, G. Curci, G. Ricciardi and A. Vicer\`e, \plb{248}{90}{181}; \\
M. Misiak, \plb{269}{91}{161}; \npb{393}{93}{23};  \\
K. Adel and Y.P. Yao, \mpla{8}{93}{1679};   \\
M. Ciuchini, E. Franco, G. Martinelli, L. Reina and L. Silvestrini,
\plb{316}{93}{127}.

\bibitem{Stefano} S. Bertolini, plenary session report at the DESY
Theory Workshop {\it Waiting for the Top Quark},
Hamburg, 1990 (unpublished); see also: in {\em Heavy
Flavours}, eds. A.J. Buras and M. Lindner (World Scientific, Singapore,
1992).

\bibitem{russi} A.A. Penin and A.A. Pivovarov, {\em Il Nuovo Cimento}
{\bf 106 A} (1993) 19.

\bibitem{matrice} M. Ciuchini \etal, in ref. \cite{qcdbsgamma}.

\bibitem{DH} J. F. Donoghue and B.R. Holstein, \prd{32}{85}1152; \\
N.G. Deshpande, X.-G. He and S. Pakvasa, preprint OITS-533 (1994).

\bibitem{model} A. Manohar and H Giorgi, \npb{234}{84}189;\\
J. Bijnens, H. Sonoda and M.B. Wise, {\em Can. J. Phys.}{\bf 64} (1986) 1;\\
D. Espriu, E. de Rafael and J. Taron, \npb{345}{90}{22}.

\bibitem{Eeg} J.O. Eeg, private communication.

\bibitem{BEF} S. Bertolini, J.O. Eeg and M. Fabbrichesi, to appear.
}

\end{thebibliography}
\end{document}